\documentstyle[12pt,epsf,rotate]{article}

\newcommand{\be}{\begin{equation}}
\newcommand{\ee}{\end{equation}}
\newcommand{\ba}{\begin{eqnarray}}
\newcommand{\ea}{\end{eqnarray}}
\newcommand{\NL}{\nonumber \\ & & \nonumber \\}

\newcommand{\LO}{\tau}
\newcommand{\grad}[1]{\frac{\partial}{\partial #1}}
\newcommand{\ddL}{ \left( \grad{\LO} + v \grad{r} \right) }
\newcommand{\ddr}{ \left( \grad{r} + v \grad{\LO} \right) }

\newcommand{\ch}{{\rm cosh}\,}
\newcommand{\sh}{{\rm sinh}\,}

\newcommand{\vs}{\vspace{0.5cm}}

\begin{document}


\baselineskip 14.5pt
\parindent=0cm
\parskip 3mm



\begin{center}
{\bf \huge 
 Analytic solution for relativistic transverse flow
 at the softest point
}

\vspace{1.0cm}

{\sc Tam\'as S. Bir\'o\footnote{e-mail: tsbiro@sunserv.kfki.hu
 ; \quad http://sgi30.rmki.kfki.hu/~tsbiro/},
}

\end{center}

\centerline{\small MTA RMKI, H-1525 Budapest, P.O.Box 49}

\vspace{0.5cm}
\centerline{PACS: 25.75}
\vs
\centerline{Keywords: transverse flow, analytic solution}


\vspace{1.0cm}
\begin{small}
{\bf Abstract:} 
We obtain an extension of Bjorken's 1+1 dimensional scaling
relativistic flow solution to relativistic transverse velocities
with cylindrical symmetry in 1+3 dimension at constant, homogeneous
pressure (vanishing sound velocity). This can be the situation
during a first order phase transition converting quark matter into
hadron matter in relativistic heavy ion collisions.

\end{small}

\vspace{0.5cm}


Hydrodynamics often allow for nonrelativistic scaling solutions.
Relativistic flow, however, seems to be an exception: besides
Bjorken's 1+1 dimensional ansatz and the spherically symmetric
relativistic expansion, no analytical solution is known \cite{Csernai}.

In this paper we present an extension of Bjorken's ansatz \cite{Bjorken}
for longitudinally and transversally relativistic
flow patterns with cylindrical symmetry in 1+3 dimensions.
This is an analytical solution of the flow equations of a perfect fluid
for physical situations when the sound velocity is zero,
$c_s^2 = dp/d\epsilon = 0$,
with energy density $\epsilon$ and pressure $p(\epsilon)$.

In particular this happens during a first order phase transition, 
the pressure is constant while the energy density changes
(in heavy ion collisons increases and drops again). This
should, in principle,  be signalled by a vanishing sound velocity.
A remnant of this effect in finite size, finite time
transitions might be a softest point of the equation of
state, where $c_s^2$ is minimal. In fact, this has been suggested
as a signal of phase transition by Shuryak\cite{Shuryak}, and investigated
numerically in several recent works \cite{Rischke, Dumitru}.

In the light of this research, the presentation of an
analytical solution including relativistic
transverse flow is worthwile. For a general equation of state
$p(\epsilon)$ the analytical solution given in this paper
cannot be extended, only a perturbative expansion in terms
of mild ($v \ll 1$) transverse velocities can be established.
Such an approximation has been recently presented in \cite{Russians}.
Nonrelativistic analytic solution has been also given several times,
with respect to heavy ions see \cite{Zim,Cs1,Cs2}.

\vspace{0.5cm}

The 1+1 dimensional Bjorken flow four-velocity is a normalized,
timelike vector. It is natural to choose this as the first of
our comoving frame basis vectors (vierbein). 
The three further, spacelike vectors
will be constructed orthogonal to this, separating the two
transverse directions. This basis fits excellently to a
cylindrical symmetry and to longitudinally extreme relativistic
flow.
\ba
e_{\mu}^0 = \left( {t \over \LO}, {z \over \LO}, 0, 0 \right), & \qquad & 
e_{\mu}^1 = \left( {z \over \LO}, {t \over \LO}, 0, 0 \right), \NL
e_{\mu}^2 = \left( 0, 0, {x \over r}, {y \over r},  \right), & \qquad &
e_{\mu}^3 = \left( 0, 0,-{y \over r}, {x \over r},  \right), 
\label{C_BASIS}
\ea

with $t$ time coordinate and $z$ longitudinal (beam-along)
coordinate, $x$ and $y$ transverse, cartesian coordinates.
The cylindrical radius is given by,
$r = \sqrt{ x^2 + y^2}$,
and $\LO$ is the {\em longitudinal proper time}:
$\LO = \sqrt{ t^2 - z^2}$.

This way the basis can be re-written in terms of 
a hyperbolic angle (coordinate-rapidity) $\eta$
and a polar angle $\phi$
\ba
e_{\mu}^0 = \left( \ch \eta, \sh \eta, 0, 0 \right), & \qquad &
e_{\mu}^1 = \left( \sh \eta, \ch \eta, 0, 0 \right), \NL
e_{\mu}^2 = \left( 0, 0,  \cos \phi, \sin \phi,  \right), & \qquad &
e_{\mu}^3 = \left( 0, 0, -\sin \phi, \cos \phi,  \right). 
\label{A_BASIS}
\ea

This basis is orthonormal,
$e_{\mu}^{a} \cdot e^{\mu, b} = g^{ab},$
and satisfies the differential relations:
\be
de_{\mu}^0 = e_{\mu}^1 d\eta, \qquad  de_{\mu}^1 =  e_{\mu}^0 d\eta, \qquad
de_{\mu}^2 = e_{\mu}^3 d\phi, \qquad  de_{\mu}^3 = -e_{\mu}^2 d\phi.
\label{DIFF_BASIS}
\ee

The space-time coordinate differential and the partial derivative
are in our basis given by,
\ba
dx_{\mu} &=& e_{\mu}^0 d\LO + e_{\mu}^1 \LO d\eta +
		e_{\mu}^2 dr + e_{\mu}^3 r d\phi, \NL
\partial_{\mu} &=& e_{\mu}^0 \frac{\partial}{\partial \LO}
    - e_{\mu}^1 \frac{1}{\LO}\frac{\partial}{\partial \eta}
    - e_{\mu}^2 \frac{\partial}{\partial r}
    - e_{\mu}^3 \frac{1}{r}\frac{\partial}{\partial \phi}.
\label{GRAD}
\ea

\vspace{0.5cm}

We consider ideal fluids (``dry water''), where the energy momentum tensor
is given by
\be
T_{\mu\nu} = (\epsilon + p) u_{\mu} u_{\nu} - p g_{\mu\nu},
\label{T_MU_NU}
\ee
and the equation of state is given in the form of $p(\epsilon)$.
The ansatz for an almost boost invariant flow with some transverse,
cylindrically symmetric component is then given by
\be
u_{\mu} = \gamma \left( e_{\mu}^0 + v e_{\mu}^2 \right),
\label{U_ANSATZ}
\ee
using the Lorentz factor $\gamma = (1-v^2)^{-1/2}$.
This four-velocity is normalized to one:
\be
u_{\mu} u^{\mu} = \gamma^2 - \gamma^2 v^2 = 1.
\ee
For the four divergence of the flow we obtain
\be
\partial_{\mu} u^{\mu} =  \grad{\LO} \gamma + \grad{r} (\gamma v)
 + \gamma  \left( \frac{1}{\LO} + \frac{v}{r} \right).
\label{FOUR_DIV}
\ee
The co-moving flow derivative is 
\be
u_{\mu} \partial^{\mu} = \gamma \ddL .
\label{CO_DER}
\ee
In order to formulate the Euler equation describing a
possible acceleretion or decceleration of the flow, one also
needs a projection orthogonal to $u_{\mu}$. We use
\be
\nabla_{\lambda} = \left( g_{\lambda \mu} - u_{\lambda} u_{\mu} \right)
\partial^{\mu},
\ee
which is given by
\be
\nabla_{\lambda} = - \gamma^2 \left( e_{\lambda}^2 + v e_{\lambda}^0 \right)
\ddr  \, - \, e_{\lambda}^1 \frac{1}{\LO} \grad{\eta}
\, - \, e_{\lambda}^3 \frac{1}{r} \grad{\phi}.
\label{TRV_DER}
\ee
The relativistic equation of local energy and momentum conservation,
$\partial_{\mu} T^{\mu\nu} = 0,$ can then be projected to a component
parallel to $u_{\mu}$ and components orthogonal to that.
Denoting the co-moving derivative (\ref{CO_DER}) by $\gamma$ times an overdot,
(i.e. $u_{\mu}\partial^{\mu} f = \gamma \dot{f}$ for any $f$), we arrive at
\ba
\gamma \dot{\epsilon} + w \partial_{\mu} u^{\mu} &=& 0, \NL
w \gamma \dot{u}_{\lambda} - \nabla_{\lambda} p  &=& 0.
\label{HYDRO}
\ea
Here we introduced the enthalpy density $w = \epsilon + p$.
The first equation of (\ref{HYDRO}) is the local form of 
the $dE+pdV=0$ adiabatic flow condition; in fact this is equivalent
to the conservation of the entropy flow in a one-component
matter.
The second equation resembles the familiar form of the
nonrelativistic Euler equation.

Summarizing we have alltogether four independent
equations,
\ba
\ddL \epsilon + w D &=& 0, \NL
\gamma^2 w \ddL v +  \ddr p & = & 0, \NL
\frac{1}{\LO} \grad{\eta} p &=& 0, \NL
\frac{1}{r} \grad{\phi} p &=& 0. 
\label{FIVE_EQS}
\ea

Here the four divergence (\ref{FOUR_DIV}) without the
Lorentz factor $\gamma$ is denoted by 
$D = \frac{1}{\gamma} \partial_{\mu} u^{\mu}$,
which can be expanded to
\be
D = \gamma^2 \ddr v  + \frac{1}{\LO} + \frac{v}{r}.
\label{DIV_FLOW}
\ee


Let us restrict ourselves now to $p=p_0={\rm constant}$ situations.
In this case $c_s^2=0$ and $w = \epsilon + p_0.$
From (\ref{FIVE_EQS}) we arrive at two independent equations only,
the first and the second one. Since derivatives of the pressure
vanish, we obtain an equation for the transverse flow component
alone:
\be
 \ddL v = 0.
\label{ANAL_FLOW}
\ee
This equation is valid even for relativistic transverse flow, it is
a quasilinear partial differential equation. A general solution
can be obtained by the factorizing ansatz $v = a(\LO)\cdot b(r)$.
The above eq.(\ref{ANAL_FLOW}) leads to
\be
\frac{a'(\LO)}{a^2(\LO)} \, + \, b'(r) \, = \, 0.
\ee
Here both terms are constant, balancing each other to zero.
The solution, which is regular at the cylindrical axis $r=0$
is given by
\be
v = \frac{\alpha r}{1 + \alpha \LO}.
\label{TRV_SOL}
\ee
The first equation of the system (\ref{FIVE_EQS}) is the
cooling equation. It reduces to
\be
 \ddL \epsilon + (\epsilon + p_0) D = 0,
\label{COOL_COOL}
\ee
with $D$ given by eq.(\ref{DIV_FLOW}).
Utilizing the solution (\ref{TRV_SOL}) for $v(\LO,r)$ we obtain
\be
D = \frac{1}{\LO} + \frac{2\alpha}{1+\alpha\LO},
\label{DIV_COOL}
\ee
which can also be written as a comoving derivative,
\be
D = \ddL \log \left( E(\LO) \right),
\label{DIV_ELO}
\ee
with
\be
E(\LO) = \LO ( 1 + \alpha \LO)^2.
\label{ELO}
\ee
Using $D$ in this form the cooling equation (\ref{COOL_COOL}) 
can be rewritten as
\be
\ddL \log \left( E(\LO)(\epsilon + p_0) \right) = 0.
\label{NICE_COOL}
\ee
A particular solution of this equation is, when the quantity
on which $\ddL$ operates is constant,
\be
\epsilon + p_0 = \frac{{\rm const.}}{\LO(1+\alpha \LO)^2}
\label{PART_SOL}
\ee
This solution interpolates between the one-dimensional cooling
law $\epsilon  \propto 1/\LO$ and a three-dimensional one
for large longitudinal proper time, $\epsilon \propto 1/\LO^3$.

There is, however, a more general solution to the ``cooling law''.
First inserting the analytic solution (\ref{TRV_SOL}) for $v$
in the co-moving derivative operation we get,
\be
\left( \frac{\partial}{\partial \LO} \, + \,
\frac{\alpha}{1+\alpha\LO} r \frac{\partial}{\partial r} \right)
\left( \log w + \log E(\LO) \right) \, = \, 0.
\ee
Multiplying this by $(1 + \alpha\LO)/\alpha$ we obtain
\be
\left( \frac{\partial}{\partial \log (1+\alpha\LO)} \, + \,
\frac{\partial}{\partial \log r} \right) 
\left( \log w + \log E(\LO) \right) \, = \, 0.
\label{TRICKY_COOL}
\ee
This equation is of type 
\be
\left( \grad{x} + \grad{y} \right) F(x,y) = 0
\label{URTYP}
\ee
whose solution is simply a function of $x-y$,
$\, F(x,y)_{{\rm solution}} = F(x-y)$.
Henceforth we obtain
\be
\log w + \log E(\LO) = F\left( \log r - \log(1+\alpha\LO) \right).
\ee
In this result one easily recognizes the analytic form of the
transverse flow velocity $v$ (\ref{TRV_SOL}) after collecting
the terms in the argument of the general function $F$ inside
one logarithm. Finally we arrive at
\be
\log w + \log E(\LO) = F(v/\alpha) = \log f(v)
\label{GEN_SOL}
\ee
as an analytic solution of the cooling law with relativistic 
cylindrical transverse flow:
\ba
\epsilon(r,\LO) &=& - p_0 + \frac{f(v)}{\LO(1+\alpha\LO)^2}, \NL
v &=& \frac{\alpha}{1+\alpha\LO} r.
\label{SOLUTION}
\ea
The unknown function of one variable $f(v)$ can be used to match
the initial profile of the radial distribution of the energy
density at $\LO=\LO_0$,
\be
\epsilon(r,\LO_0) = - p_0 + \frac{1}{\LO_0(1+\alpha\LO_0)^2} 
f \left( \frac{\alpha r}{1+\alpha\LO_0}\right).
\label{INIT}
\ee

\vspace{0.5cm}

Let us finally discuss some properties of this analytic
solution. First one would like to be convinced that
the Bjorken scaling limit is contained in eq.(\ref{SOLUTION}).
(The 1+1 dimensional solution was presented by Gyulassy and
Matsui in 1984 \cite{GyuMa84}.)
This is indeed the case, for $v \rightarrow 0$ one obtains
$\alpha = 0$ and
\be
\epsilon(r,\LO)  = - p_0 + \frac{f(0)}{\LO},
\ee
leading to
\be
\left( \epsilon(\LO) + p_0 \right) = (\epsilon_0+p_0) \frac{\LO_0}{\LO}.
\label{BJORKEN}
\ee

As an example of the $p = {\rm const}.$ situation let us consider
an oversimplified system: massless quark-gluon plasma described
by the bag equation of state,
\be
\epsilon_Q = \sigma T^4 + B, \qquad
p_Q = \frac{1}{3} \sigma T^4 - B,
\label{BAG_EOS}
\ee
keeps a Gibbs equilibrium with a light, relativistic pion gas,
described by
\be
\epsilon_H = h T^4, \qquad
p_H = \frac{1}{3} h T^4.
\label{PION_EOS}
\ee
During a first order phase transition, for simplicity assumed
to take place in the total volume, the pressure is constant
and homogeneous (Gibbs criterion). Actually from this requirement
$p_Q = p_H$ one usually obtains the transition temperature
due to
\be
\frac{1}{3} (\sigma - h) T_c^4 = B.
\label{TRANS}
\ee
The energy density is that of a mixture, containing $x$ part
quark matter and $(1-x)$ part hadron matter,
\be
\epsilon = x \epsilon_Q(T_c) + (1-x) \epsilon_H(T_c).
\ee
This partition $x$ is what changes according to the cooling law,
respectively its solution (\ref{SOLUTION}).
Utilizing eqs.(\ref{BAG_EOS},\ref{PION_EOS}) and (\ref{TRANS})
we arrive at
\be
\epsilon = 4B x(\LO) + \epsilon_H = -p_H + (\epsilon_Q + p_Q)
\frac{\LO_0}{\LO}.
\ee
Assuming at $\LO=\LO_0$ pure quark matter $x(\LO_0)=1$, the time
of the total conversion $\LO_1$ when $x(\LO_1)=0$ is given by
\be
\LO_1 = \frac{\epsilon_Q + p_Q}{\epsilon_H + p_H} \LO_0
= \frac{\sigma}{h} \LO_0.
\ee
It is determined by the relative number of degrees of freedom
in the two phases in this simple scenario.

The same discussion is somewhat more complex using the analytic
solution with a cylindrical flow. At the beginning of the
phase transition $x(\LO_0)=1$ let be a transverse velocity,
\be
v(\LO_0, r) = v_0 \frac{r}{R_0}.
\ee
The analytic solution (\ref{TRV_SOL}) leads to
\be
v(\LO, r) = \frac{v_0}{R_0 + v_0 (\LO - \LO_0)} r
\label{TRV_SLOW_DOWN}
\ee
featuring a slowing down of the transverse flow. From the initial condition
for the transverse flow we also obtained the parameter $\alpha$
\be
\alpha = \frac{v_0}{R_0 - v_0\LO_0}.
\ee
Since the four-flow is given by
$$ u_{\mu} = \gamma \left( e_{\mu}^0 + v e_{\mu}^2 \right)$$
a slowing down of the transverse component implies a slowing
down of the original Bjorken component as well ($\gamma$ is decreasing
to one, when $v$ is decreasing to $0$).

The cooling of the energy density can be obtained now by fitting
the profile at $\LO=\LO_0$ to $\epsilon(r,\LO_0) = e(r),$
\be
\epsilon(r,\LO) + p_0 \, = \,  \frac{\LO_0}{\LO} \, \frac{R_0^2}{R^2} 
\left( e\left( \frac{R_0}{R} \, r \, \right)
\, + p_0 \, \right)
\label{ENERGY_TRV}
\ee
with
\be
R(\LO) = R_0 + v_0(\LO - \LO_0).
\ee

This solution follows the Bjorken cooling law at the beginning
$\LO \approx \LO_0$,
\be
\epsilon(r,\LO) + p_0 \approx \frac{\LO_0}{\LO} \left( e(r) + p_0 \right),
\ee
but for long times $\LO \gg \LO_0$ turns over to a
three-dimensional scaling
\be
\epsilon(r,\LO) + p_0 \approx \frac{\LO_0}{\LO^3} \frac{R_0^2}{v_0^2}
\left( e\left(\frac{R_0}{v_0\LO}r\right) + p_0 \right).
\ee
Here for large times, the driving force besides the constant pressure
will be $e(0)$, the original energy density at the axis.


Finally we repeat the calculation in the simplified scenario
assuming a first order phase transition between ideal quark gluon
plasma and ideal pion gas. The general solution for the
quark matter part, $x$ is given by
\be
x(\LO,r) = - \frac{h}{\sigma -h} +
\frac{\LO_0 R_0^2}{\LO R^2} \left(
\frac{h}{\sigma - h} + x(\LO_0, \frac{R_0}{R} r) \right).
\ee
Assuming a linearly decreasing transverse profile of the QGP part initially
(at the beginning of the phase transition),
\be
x(\LO_0,r) = 1 - r / R_1,
\ee
we obtain the transverse radius of the mixed phase as
\be
r_b = \frac{R_1}{R_0} R \left( 1 - \frac{h}{\sigma -h}
\left(\frac{\LO R^2}{\LO_0 R_0^2} - 1 \right) \right).
\ee
In the simple case, when $R_0 = v_0 \LO_0$, the scaling
transverse radius $R$ is proportional to the longitudinal
proper time $\LO$: $R=v_0 \LO$. The transverse velocity
field is also particularly simple: $v = r/\LO$.
The above result simplifies to
\be
r_b = R_1 \frac{\LO}{\LO_0} \left( \frac{\sigma}{\sigma - h}
- \frac{h}{\sigma - h} \frac{\LO^3}{\LO_0^3} \right).
\ee
This expression initially grows, achieves a maximum and then
decreases towards zero. The maximum,
\be
r_b^{{\rm max}} = \frac{3}{4} R_1 \left( \frac{\sigma}{4h} \right)^{1/3}
\frac{\sigma}{\sigma - h},
\ee
is achieved at time
\be
\LO^{{\rm max}} = \LO_0 \left( \frac{\sigma}{4h} \right)^{1/3},
\ee
the $x=0$ for the whole space (equivalent to $r_b=0$) at
\be
\LO_1 = \LO_0 \left( \frac{\sigma}{h} \right)^{1/3}.
\ee
Realistic estimates use $\sigma = 37$ for the quark gluon plasma
and $h = 3$ for the pion gas. This leads to
$r_b^{{\rm max}} \approx 1.14 R_1$, 
$\LO^{{\rm max}} \approx 1.45 \LO_0$.
and $\LO_1 \approx 2.31 \LO_0$. It is realistic to assume 
$\LO_0 \approx 5$ fm/c for a CERN SPS
Pb+Pb experiment. The radial extension of the quark matter
can grow with about 14 per cent and then the conversion into
hadrons eats it up, reaching the zero radius in about 11.5 fm/c.
This is a fast hadronization even in this simple scenario.

For a smaller initial radial flow, realistically $v_0 = 0.6$
at the radial edge of the cylinder $r=R_0$, one obtains
$v_0\LO_0/R_0 = 0.428$. Fig.1 shows the radial space - time
evolution profiles of the mixture: the outmost curve corresponds
to $x = 0$ (pure hadron matter) and the steps are $0.1$. The innermost
curve belonging to $x=1.0$ shrinks to a point, since the
phase conversion starts immediately at $\tau = \tau_0$.
Here we assumed a linear initial profile $x(r,\LO_0) = 1 - r/R_0.$

With these parameters the mixed phase of $R_0 = 7$ fm longs for
about $\LO_b \approx 17.5$ fm/c.

\vs
In conclusion we have presented an analytical solution
for relativistic transverse flow of a perfect fluid
at the softest point. This can be realized during a
first order phase transition, when the pressure is
homogeneous and constant in a large volume. The phase
conversion during this stage of expansion is given by
this analytical solution which scales with the
longitudinal proper time initially like the Bjorken
flow $1/\LO$, but eventually like a spherical flow
$1/\LO^3$. For simple equations of state for the quark and
hadronic matter side the estimated longitudinal proper time
spent in the mixed phase turns out to be more than twice
the time when the phase transition began.

\vspace{1.0cm}

Acknowledgements: This work is part of a collaboration between the
Deutsche Forschungsgemeinschaft and the Hungarian Academy of
Science (project No. DFG-MTA 101/1998) and has been
supported by the Hungarian National Fund for Scientific
Research OTKA (project No. T019700 and T024094),
as well as by the Joint Hungarian-American Technology Fund
TeT-MAKA (no. JF.649).

Discussions with T.~Cs\"org\H{o}, L.~P.~Csernai, S.~G.~Matinyan, B.~M\"uller,  
and J.~Zim\'anyi are gratefully acknowledged.

\vspace{0.5cm}



\begin{figure}
\vskip -0.4in
\epsfxsize=3.5in
\epsfysize=3.5in


\centerline{\rotate[r]{\epsffile{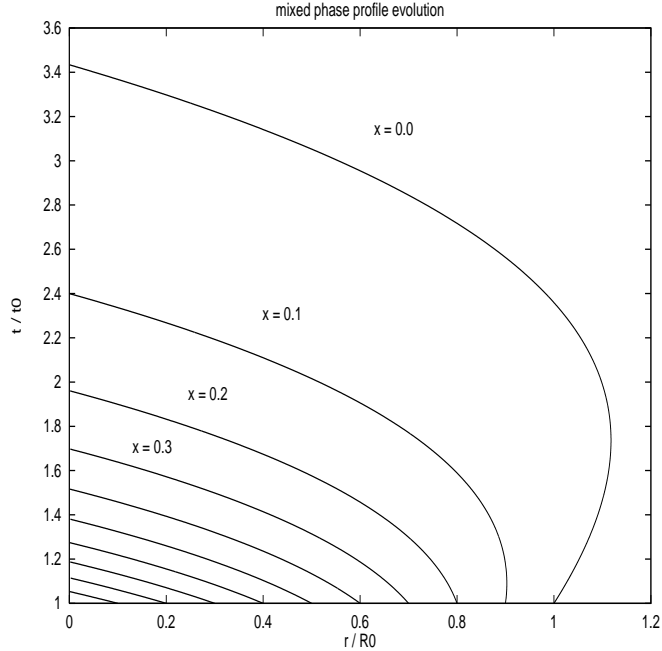}}}

\vskip -0.05in
\caption[]{
 \label{figure1}
Contour lines of several QGP - hadronic mixtures
(from 0 to 1 by 0.1 steps) in the $\tau/\tau_0$ - $r/R_0$
plane. The initial profile at $\tau = \tau_0$ was assumed
to be linear giving zero percent quark matter exactly at
$R_1 = R_0$. The initial transverse flow is $v_0 = 0.6$
at the edge $R=R_0$.

}
\end{figure}

\vs



\begin{thebibliography}{lblxxxxx}

\bibitem{Csernai} L.~P.~Csernai: {\em Introduction to Relativistic
	Heavy Ion Collisions,} John Wiley \& Son, Chicester, 1994,
	L.~P.~csernai, Heavy Ion Physics 5, 321, 1997.

\bibitem{Bjorken} J.~D.~Bjorken, Phys. Rev. D27, 140, 1983.

\bibitem{Shuryak} C.~M.~Hung, E.~V.~Shuryak, Phys.Rev.Lett. 75, 4003, 1995

\bibitem{Rischke} 
   D.~H.~Rischke: {\em Fluid dynamics for relativistic nuclear collisions,}
   nucl-th/9809044,

        J.~Brachmann, S.~Soff, A.~Dumitru, H.~St\"ocker,
	J.~A.~Maruhn, W.~Greiner, D.~H.~Rischke: 
   {\em Antiflow of nucleons at the softest point of EOS}, nucl-th/9908010,

\bibitem{Dumitru} A.~Dumitru, D.~H.~Rischke, Phys. Rev. C 59, 354, 1999

\bibitem{Russians} P.~Milyutin, N.~Nikolaev, Heavy Ion Physics 8, 333, 1998

\bibitem{Zim}	J.~P.~Bondorf, S.~I.~A.~Garpman, J.~Zim\'anyi,
	Nucl. Phys. A 296, 320, 1978

\bibitem{Cs1} P.~Csizmadia, T.~Cs\"org\H{o}, B.~Luk\'acs,
	Phys. Lett. B 443, 21, 1998

\bibitem{Cs2} T.~Cs\"org\H{o}: {\em Simple analytic solution of fireball
	hydrodynamics}, nucl-th/9809011

\bibitem{GyuMa84} M.~Gyulassy, T.~Matsui, Phys.Rev.D 29, 419, 1984


\end{thebibliography}
\end{document}